\documentclass[preprint,showpacs,preprintnumbers,amsmath,amssymb]{revtex4}
\usepackage{amsmath,amssymb,graphics,epsfig,subfigure}
\usepackage{color}

\begin{document}
\newcommand {\nn} {\nonumber}
\renewcommand{\baselinestretch}{1.3}

\title{Equatorial and quasi-equatorial gravitational lensing by Kerr black hole pierced by a cosmic string}
\author{Shao-Wen Wei$^{1,\;2}$ \footnote{weishw@lzu.edu.cn},
        Yu-Xiao Liu$^{1}$ \footnote{liuyx@lzu.edu.cn, corresponding author}}.

\affiliation{$^{1}$ Institute of Theoretical Physics, Lanzhou University, Lanzhou 730000, People's Republic of China\\
$^2$ Interdisciplinary Center for Theoretical Study,
University of Science and Technology of China, Hefei, Anhui 230026,
People's Republic of China}

\preprint{USTC-ICTS-11-09}

\begin{abstract}
In the present paper, we study numerically the equatorial lensing and quasiequatorial lensing by Kerr black hole pierced by a cosmic string in the strong deflection limit. We calculate the strong deflection limit coefficients and the deflection angle, which are found to depend closely on the cosmic string parameter $\beta$ and dimensionless spin $a_{*}$. The magnification and positions of relativistic images are also computed in the strong deflection limit and a two-dimensional lens equation is derived. The most important and outstanding effect is that the caustics drift away from the optical axis and shift in the clockwise direction with respect to the Kerr black hole. For fixed $a_{*}$ of the black hole, the caustics drift farther away from the optical axis for a large value of $\beta$. And for fixed $\beta$, they drift farther for high $a_{*}$. We also obtain the intersections of the critical curves with the equatorial plane, which decrease with $a_{*}$ and $\beta$. In particular, we obtain a quantity $\bar{\mu}_{k+1}/\bar{\mu}_{k}$, which is independent of the black hole spin and mass. Thus, through measuring it, one is allowed to determine the value of $\beta$ from astronomical observations.
\end{abstract}



\maketitle

\section{Introduction}

One of the robust tests of general relativity is the bend of light passing the Sun. The phenomenon is referred to as gravitational lensing, and the object causing a detectable deflection is known as gravitational lens. On one side, with the information of the lens, we can estimate the deflection angle. On the other side, with the deflection angle from astronomical observation, we can determine the nature of the lens. The latter property can be used to probe the nature of distant stars and galaxies, or even the nature of black holes. It also provides a profound way to determine the cosmological parameters \cite{Sutherland}, as well as a way to verify alternative theories of gravity \cite{Bekenstein,Eiroa,Sarkar,Chen1}. Furthermore, it can guide us to detect the gravitational waves at proper frequency \cite{Stefanov,WeiLiu}.

Recently, gravitational lensing has received much attention,
mainly due to the strong evidence about the presence of supermassive black holes at the centers of galaxies. The study of the gravitational lens by black holes and compact objects can be simplified by using the strong deflection limit, which can be traced back to \cite{Darwin}. The author found that the photons passing very close to a black hole could make one or more complete loops around the black hole before reaching the observer. Then, besides the two classical weak field images, there would be two infinite series of images close to the black hole. The results were rediscovered by \cite{Luminet}. On the other hand, the lens equations were studied in \cite{Frittelli,Virbhadra}, where the authors proposed a definition of an exact lens equation without reference to the black hole background, which was then applied to the Schwarzschild black hole. The results were also extended to the naked singularities lensing \cite{Virbhadra2}, Reissner-Nordstrom black hole lensing \cite{EiroaRomero}, and spherically symmetric and static black hole lensing \cite{Perlick}.

Based on the Virbhadra-Ellis lens equation, Bozza \emph{et al}. \cite{BozzaCapozziello} proposed a new and reliable method to obtain the deflection angle in the strong field limit for a Schwarzschild black hole lensing, where only the first two leading order terms were retained. Adopting the approximation, a fully analytical treatment was developed. The results showed that the deflection angle diverges logarithmically as the light rays get close to the photon sphere of the black hole. The magnification and positions of images were also found in simple expressions. Subsequently, Bozza \cite{Bozza} generalized the results to an arbitrary spherically symmetric spacetime. The treatment was also applied to other black holes, naked singularities and wormholes \cite{Bozza72,Bozza74,Eiroa66,Eiroa71,Whisker,Bhadra,Vazquez,
Eiroa69,Nandi,Bozza76,Virbhadra77,Virbhadra79,Bisnovatyi,Yazadjiev,
Nun,Liu,Ghosh,Chengc28,Chengc29,Cheng27,Eiroa1011,Ding,Chen1102,Wei,DingJing}.

In general, the best candidate for strong field gravitational lensing is the one in the center of a galaxy. Motivated by the idea that these black holes have spin, Bozza \cite{Bozza67}, and Gyulchev and Yazadjiev \cite{Gyulchev75} studied quasiequatorial gravitational lensing by rotating black holes, where the trajectory of the photons is allowed to have a small inclination. They showed that, a rotating black hole has a spin-dependent photon sphere. The deflection angle, positions and magnification of these images were all found to depend on the spin of the rotating black hole.

On the other hand, the study of space-times containing cosmic strings has gained a lot of renewed interest in recent years. In particular, the complete set of solutions of the geodesic equations for a particle in the Schwarzschild black hole and Kerr black hole pierced by a cosmic string were studied in \cite{Galtsov,Hackmann0,Hackmann}. The result shows that the cosmic string parameter has a significant influence on the geodesic equations and light deflection \cite{Hackmann}. Its gravitational effect can also be found in \cite{Aliev}. Since the black hole lensing may provide us with information on the lens, such as the spin, charge, gravitomagnetic mass and so on, we would like to study the equatorial and quasiequatorial gravitational lensing by a Kerr black hole pierced by a cosmic string in the strong field limit. Besides the spin, we also expect to study the influence of the cosmic string on black hole lensing. This may provide us with a robust test to probe the cosmic string in our Universe.

The paper is structured as follows. In Sec. \ref{geodesic}, we give a brief review of the null geodesics for the Kerr black hole pierced by a cosmic string. In Sec. \ref{Equatorial}, we calculate numerically the strong deflection limit coefficients and the strong deflection angle. In Sec. \ref{Quasiequatiorial}, we investigate quasiequatorial lensing by the Kerr black hole pierced by a cosmic string. The magnification and caustic points are discussed in Sec. \ref{Magnification}. In Sec. \ref{Critical}, the critical curves and caustic structures are considered. The final section is devoted to a brief discussion.

\section{Kerr black hole pierced by a cosmic string and null geodesics}
\label{geodesic}

In this section, we would like to give a brief introduction to the Kerr black hole pierced by a cosmic string and the null geodesics in spacetime. The line element is, in generalized Boyer-Lindquist coordinates,
\begin{eqnarray}
 ds^{2}=&-&\bigg(1-\frac{2M x}{\rho^{2}}\bigg)dt^{2}
         +\frac{\rho^{2}}{\Delta}dx^{2}+\rho^{2}d\vartheta^{2}\nonumber\\
         &+&\beta^{2}\bigg(x^{2}+a^{2}
            +\frac{2Mx a^{2}\sin^{2}\vartheta}{\rho^{2}}\bigg)\sin^{2}\vartheta
                         d\phi^{2}
         -\beta\frac{4Mxa\sin^{2}\vartheta}{\rho^{2}}dtd\phi, \label{metric}
\end{eqnarray}
where
\begin{eqnarray}
 \Delta  &=& x^{2}-2Mx+a^{2},\\
 \rho^{2}&=& x^{2}+a^{2}\cos^{2}\vartheta.
\end{eqnarray}
The Komar mass and angular momentum are \cite{Galtsov}
\begin{eqnarray}
 M_{\text{phys}}&=&\frac{1}{8\pi}\oint K_{(t)}^{\mu;\nu}d\Sigma_{\mu\nu}
                =M\beta,\\
 J_{\text{phys}}&=&\frac{1}{16\pi}\oint K_{(\phi)}^{\mu;\nu}d\Sigma_{\mu\nu}
                =J\beta,
\end{eqnarray}
where $K_{(t)}=\partial/\partial t$ and $K_{(\phi)}=\beta^{-1}\partial/\partial\phi$ are two commuting Killing vector fields. Thus the spin parameter $a_{\text{phys}}=J_{\text{phys}}/M_{\text{phys}}=J/M=a$ remains unchanged. The cosmic string parameter $\beta$ is related to the deficit angle as $\delta=2\pi(1-\beta)$ with $0<\beta\leq 1$. The metric (\ref{metric}) describes a Kerr black hole pierced by an infinitely thin cosmic string aligned with the rotation axis of the black hole, $\vartheta=0,\;\pi$. The deficit angle can be written in terms of the energy per unit length $l$ of the cosmic string: $\delta=8\pi G_{4}l$. Note that, for $a=0$ and $\beta\neq 1$, the metric (\ref{metric}) describes a Schwarzschild black hole pierced by an infinitely thin cosmic string \cite{Aryal}. For $\beta=1$, it reduces to the Kerr black hole metric, and for $\beta=1$ and $a=0$, it describes a Schwarzschild black hole.

The horizon of this black hole is determined by $\Delta=0$, which gives
\begin{eqnarray}
 x_{\pm}=M\pm\sqrt{M^{2}-a^{2}}.
\end{eqnarray}
A black hole solution needs $x_{+}\geq x_{-}$ or $M^{2}\geq a^{2}$ ($M^{2}_{\text{phys}}\geq \beta^{2}a^{2}$), where the inequality is saturated for the extremal black hole. The photon will be absorbed by the black hole if it crosses the outer horizon. Therefore, we just consider the case $x>x_{+}$.

Measuring the distances in Schwarzschild radius ($r_{\text{s}}=2M_{\text{phys}}$), a dimensionless metric of (\ref{metric}) can be obtained by defining
\begin{eqnarray}
 \tilde{s}=\frac{s}{r_{\text{s}}},\quad
 r=\frac{x}{r_{\text{s}}},\quad
 T=\frac{t}{r_{\text{s}}},\quad
 a_{*}=\frac{a}{r_{\text{s}}},
\end{eqnarray}
which is equivalent to setting $2M_{\text{phys}}=1$. In the following, we will adopt this dimensionless metric for convenience. In this case, it describes the Kerr black hole when $\beta=1$, and the Schwarzschild black hole when $\beta=1$ with dimensionless spin $a_{*}=0$.

The geodesic equations for a test particle in the background spacetime (\ref{metric}) and its full set of solutions have been obtained by Hackmann \emph{et al.} \cite{Hackmann}. Here we would like to give some notes on the conical singularity. As we know, there is a deficit angle $\delta=2\pi(1-\beta)$ in this black hole background. So the equatorial plane is a cone with the azimuthal angle's period $2\pi\beta$, if cutting it open and flatting it in a flat plane, there will be a deficit angle $\delta$. However, one can introduce a rescaling azimuthal angle $\phi$ such that it has period $2\pi$. In this case, if one wants to measure an arbitrary azimuthal angle on the cone, a multiplicative factor $\beta$ will be enough. In the following, we will measure all the azimuthal angles on the cone, which means all angles are measured by $\beta\phi$. For a photon traveling in spacetime, the geodesic equations read
\begin{eqnarray}
 \tilde{\rho}^{2} \dot{r}&=&\pm \sqrt{\mathcal{R}},\label{rt}\\
 \tilde{\rho}^{2} \dot{\vartheta}&=&\pm\sqrt{\Theta},\label{Theta}\\
 \tilde{\rho}^{2} \beta\dot{\phi}&=&L\sin^{-2}\vartheta-a_{*}E
           +\frac{a_{*}}{\tilde{\Delta}}\bigg((r^{2}+a_{*}^{2})E-a_{*}L\bigg), \label{phit}\\
 \tilde{\rho}^{2} \dot{T}&=&\frac{E(r^{2}+a_{*}^{2})^{2}-ra_{*}L/\beta}{\tilde{\Delta}}-a_{*}^{2}E\sin^{2}\vartheta,\label{tt}
\end{eqnarray}
where the dot indicates the derivative with respect to the affine parameter. $\mathcal{R}$ and $\Theta$ are only functions of $r$ and $\vartheta$, respectively, and they are
\begin{eqnarray}
 \mathcal{R} &=&(a_{*}L-(r^{2}+a_{*}^{2})E)^{2}-\tilde{\Delta}((L-a_{*}E)^{2}+\mathcal{K}),\\
 \Theta      &=&\mathcal{K}-\cot^{2}\vartheta(L^{2}-a_{*}^{2}E^{2}\sin^{2}\vartheta),
\end{eqnarray}
where $\tilde{\rho}^{2}=r^{2}+a_{*}^{2}\cos^{2}\theta$, $\tilde{\Delta}=
r^{2}-r/\beta+a_{*}^{2}$, and $\mathcal{K}$ is a separation constant of motion, known as the Carter constant. Two other new parameters, $E$ and $L$, are the conservation of energy and orbital angular momentum per unit mass of the motion, and they correspond to the Killing fields $K_{(t)}$ and $K_{(\phi)}$, respectively. Note that the geodesic equations closely depends on the cosmic string parameter $\beta$. Thus, we can conclude that the parameter $\beta$ should influence the motion of the photon, and it is natural to conjecture that there is a $\beta$-dependent gravitational lensing. Moreover, we can express the lightlike geodesics, in terms of integrals, as
\begin{eqnarray}
 \int^{r}\frac{dr}{\pm \sqrt{\mathcal{R}}}
         =\int^{\vartheta}\frac{d\vartheta}{\pm \sqrt{\Theta}},
\end{eqnarray}
\begin{eqnarray}
 \beta\Delta\phi=a_{*}\int^{r}\frac{(r^{2}+a_{*}^{2})E-a_{*}L}{\pm\tilde{\Delta}\sqrt{\mathcal{R}}}dr
             +\int^{\vartheta}\frac{L\sin^{-2}\vartheta-a_{*}E}{\pm\sqrt{\Theta}}
                d\vartheta.
\end{eqnarray}
Positive signs of $\sqrt{\mathcal{R}}$ and $\sqrt{\Theta}$ correspond to the case where the lower integration limit is smaller than the upper limit, and negative ones correspond to the opposite case.

We suppose here that the gravitational field far away from the black hole is very weak and can be described by a flat metric. Then we can assume that the light ray trajectory is a straight line at infinity and it bends near the black hole. From this viewpoint, we can identify the approximate light ray with three parameters, $\psi_{o}$, $u$, and $h$. The first parameter, $\psi_{o}$, denotes the inclined angle that the incoming light ray forms with the equatorial plane. The second one, $u$, is an impact parameter of the projection of the light ray trajectory in the equatorial plane. And the last one, $h$, denotes the distance between that point of the projection which is closer to the black hole and the trajectory itself. The schematic illustration can be found in Fig. 1 of \cite{Bozza67}.

As suggested in \cite{Vazquez,Bozza67,Gyulchev75}, if the observer is located at coordinates $(r_{o},\;\vartheta_{o})$ in the Boyer-Lindquist system, we then can define two celestial coordinates $\zeta_{1}$ and $\zeta_{2}$ for an image. The coordinate $\zeta_{1}$ measures the observable distance of the image with respect to the symmetry axis in the direction normal to the line of sight, and the coordinate $\zeta_{2}$ describes the observable distance from the image to the source projection in the equatorial plane in the direction orthogonal to the line of sight. With the help of Eqs. (\ref{rt})-(\ref{tt}), we can express the two coordinates $\zeta_{1}$ and $\zeta_{2}$ as
\begin{eqnarray}
 \zeta_{1}&=&r_{o}^{2}\sin\vartheta_{o}\frac{\beta d\phi}{dr}\bigg|_{r,\;r_{o}\rightarrow\infty}
           =L\sin^{-1}\vartheta_{o},\\
 \zeta_{2}&=&r_{o}^{2}\frac{d\vartheta}{dr}\bigg|_{r,\;r_{o}\rightarrow \infty}
           =h\sin\vartheta_{o},
\end{eqnarray}
where we have set $E=1$. Considering the quasiequatorial case (i.e., $\vartheta_{o}=\pi/2-\psi_{o}$ with $\psi_{o}$ a very small
value) and $\zeta_{1}=u$, we can obtain the orbital angular momentum $L$ and Carter constant $\mathcal{K}$, in terms of $u$ and $\psi_{o}$:
\begin{eqnarray}
 L          &=&u\cos\psi_{o},\label{angular}\\
 \mathcal{K}&=&h^{2}\cos^{2}\psi_{o}+(u^{2}-a_{*}^{2})\sin^{2}\psi_{o}. \label{cater}
\end{eqnarray}
Note that the expression of the Carter constant $\mathcal{K}$ (\ref{cater}) is the same as that for the Kerr black hole without cosmic string.

\section{Equatorial lensing by A Kerr black hole pierced by a cosmic string}
\label{Equatorial}

In this section, we would like to study the deflection angle of the photon and the relativistic images for black hole lensing in the equatorial plane. This means that both the observer and the source lie in the equatorial plane, and the whole trajectory of the photon is also limited in the plane.

\subsection{Reduced metric and effective potential}
\label{Reduced}

\begin{figure}
\includegraphics[width=8cm]{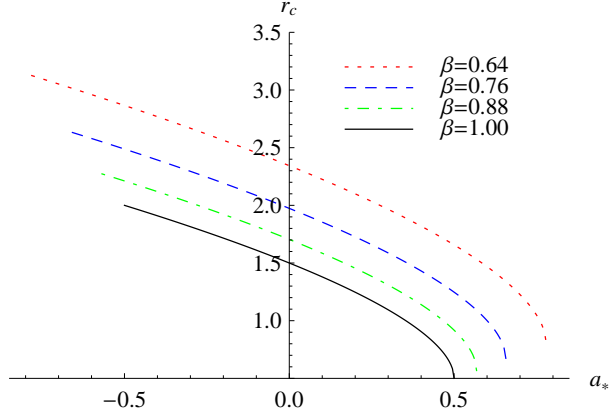}
\caption{Behaviors of the circular radius $r_{\text{c}}$ with $\theta=\pi/2$.} \label{RH}
\end{figure}

Let us start with the equatorial plane conditions $\vartheta=\frac{\pi}{2}$ and $h=\psi_{0}=0$. The reduced metric can be expressed as
\begin{eqnarray}
 d\tilde{s}^{2}=-A(r)dT^{2}+B(r)dr^{2}+C(r)d\phi^{2}-D(r)dTd\phi, \label{newmetric}
\end{eqnarray}
with the metric coefficients given by
\begin{eqnarray}
 A(r)&=&1-\frac{1}{\beta r},\\
 B(r)&=&\frac{\beta r^{2}}{\beta (r^{2}+a_{*}^{2})-r},\\
 C(r)&=&\beta\frac{\beta r^{3}+a_{*}^{2}(1+\beta r)}{r},\\
 D(r)&=&2\frac{a_{*}}{r}.\label{metric2}
\end{eqnarray}
In the equatorial plane, the first-order geodesic equations (\ref{rt})-(\ref{tt}) for the photon can be rewritten, in terms of the metric coefficients $A(r)$, $B(r)$, $C(r)$, and $D(r)$, as
\begin{eqnarray}
 \dot{T}&=&\frac{4CE-2\beta DL}{4AC+D^{2}},\label{ttdot}\\
 \dot{r}&=&\pm 2\sqrt{\frac{CE^{2}-\beta DEL-\beta^{2} AL^{2}}
        {B(4AC+D^{2})}},\label{rrdot}\\
 \dot{\vartheta}&=&0,\\
 \dot{\phi}&=&\frac{2DE+4\beta AL}{4AC+D^{2}}.\label{ppdot}
\end{eqnarray}
Without loss of generality, we take the choice $E=1$. We express (\ref{rrdot}) in the form
\begin{eqnarray}
 \bigg(\frac{dr}{d\tau}\bigg)^{2}+V_{\text{eff}}=0,
\end{eqnarray}
with the effective potential given by
\begin{eqnarray}
 V_{\text{eff}}&=&-4\frac{C-\beta DL-\beta^{2} AL^{2}}{B(4AC+D^{2})}\nonumber\\
               &=&\frac{(r-1)L^{2}}{r^{3}}+\frac{2a_{*}L}{r^{3}}
                   -\frac{a_{*}^{2}(r+1)}{r^{3}}-1.
\label{veff}
\end{eqnarray}
Now, let us turn to the photon sphere. For a spherically symmetric and static black hole spacetime, the definition of the photon sphere can be found in \cite{Virbhadra,Claudel}. However, for the spin $a\neq 0$, there is no such photon sphere. Actually, we could obtain the circular radius of the light ray for the reduced metric (\ref{newmetric}) from the effective potential $V_{\text{eff}}$ in the equatorial plane. The radius of the photon circle is consistent with that of the photon sphere when $a_{*}\rightarrow 0$ \cite{Perlick2}. The radius of the photon circle satisfies the following conditions
\begin{eqnarray}
 &&V_{\text{eff}}=0, \label{condition1}\\
 &&V_{\text{eff}}'=0, \label{condition2}
\end{eqnarray}
where the prime indicates the derivative with respect to $r$. Supposing that the photon is at the minimum distance $r_{0}$ of its trajectory, we have $\dot{r}|_{r=r_{0}}=0$ and (\ref{condition1}) becomes
\begin{eqnarray}
 \beta^{2} L^{2}A_{0}+\beta LD_{0}-C_{0}=0.\label{result1}
\end{eqnarray}
From this, we obtain the angular momentum $L$,
\begin{eqnarray}
 L=u&=&\frac{-D_{0}+\sqrt{4A_{0}C_{0}+D_{0}^{2}}}{2A_{0}\beta} \nonumber\\
    &=&\frac{-a_{*}+r_{0}\sqrt{\beta(\beta r_{0}^{2}-\beta r_{0}+a_{*}^{2})}}
                 {\beta r_{0}-1}.\label{Luangular}
\end{eqnarray}
The subscript ``0" denotes that the metric coefficients are evaluated at $r_{0}$. Here, we have set the sign before the square root to be positive, which corresponds to the light ray winding counterclockwise. Therefore, the black hole and the photon rotate in the same direction for $a_{*}>0$ and in the converse direction for $a_{*}<0$. The second condition (\ref{condition2}) is, at the minimum distance $r_{0}$,
\begin{eqnarray}
 \beta^{2} L_{0}^{2}A_{0}'+\beta L_{0}D_{0}'-C_{0}'=0.
\end{eqnarray}
With the help of (\ref{result1}), we have the following photon circle equation
\begin{eqnarray}
 A_{0}C_{0}'-A_{0}'C_{0}+\beta L(A_{0}'D_{0}-A_{0}D_{0}')=0.
\end{eqnarray}
For vanishing $D(r)$, it is just the photon sphere equation for a nonrotating black hole. Solving the equation, we will obtain the circular radius $r_{\text{c}}$. For the metric (\ref{newmetric}), the photon circle equation takes the form
\begin{eqnarray}
 r_{\text{c}}(2\beta r_{\text{c}}-3)^{2}-8\beta a_{*}^{2}=0.
\end{eqnarray}
It is clear that the cosmic string parameter $\beta$ has a significant influence on the circular radius $r_{\text{c}}$. From Fig. \ref{RH}, we can see that, for the fixed $\beta$, direct photons (winding in the same direction of rotation as the black hole) always have a smaller circular radius than retrograde ones (winding in the converse direction of rotation as the black hole). Thus, one can get the result that retrograde photons are captured more easily than direct ones. For fixed $a_{*}$, the circular radius decreases with $\beta$, which means that the photons are captured more easily by a black hole with a small cosmic string parameter $\beta$.

For the spherically symmetric and static black hole, one of the important physical properties of its photon sphere is that, when $r_{0}$ approaches it, the deflection angle will be unlimited. We will show in the next section that this property also holds for the circular radius. On the other hand, for the case $\beta=1$, we will obtain the radius of the photon circle for a Kerr black hole. For the Schwarzschild black hole, we get $r_{\text{c}}=\frac{3}{2}$, which can be also found in Fig. \ref{RH} as shown by the black solid line at $a_{*}=0$.

\subsection{Deflection angle in the equatorial plane}
\label{Deflectionequatorial}

In this subsection, we would like to study black hole lensing in the equatorial plane in the strong deflection limit.

With the help of (\ref{rt}) and (\ref{phit}), we find the azimuthal shift as a function of the distance,
\begin{eqnarray}
 \beta\frac{d\phi}{dr}=\frac{\sqrt{B|A_{0}|}(D+2\beta LA)}
     {\sqrt{4AC+D^{2}}\sqrt{\text{sgn}(A_{0})\big[CA_{0}-AC_{0}
       +\beta L(AD_{0}-A_{0}D)\big]}}, \label{azimuthal}
\end{eqnarray}
where sgn($X$) gives the sign of $X$. Suppose that the azimuthal angles for the departure and approach of the photon are the same; then the whole deflection angle can be expressed via the integration of (\ref{azimuthal}) from $r_{0}$ to infinity,
\begin{eqnarray}
 \alpha(r_{0})=\phi_{f}(r_{0})-\pi\beta,\label{deflection}
\end{eqnarray}
It is worth noting that we have considered the influence of conical singularity. The total azimuthal angle is given by
\begin{eqnarray}
 \phi_{f}(r_{0})=2\int^{\infty}_{r_{0}}\beta\frac{d\phi}{dr}dr.\label{azimuthal11}
\end{eqnarray}
For weak gravitational field lensing, we have a very small deflection angle. However, for a black hole lensing, the deflection angle is allowed to take a large value. With a detailed examination, one could find that the total azimuthal angle $\phi_{f}(r_{0})$ monotonically increases with as $r_{0}$ decreases, and it will be unlimited as $r_{0}\rightarrow r_{\text{c}}$. So, this implies that a photon could make one or more complete loops around the black hole before reaching the observer.

Next, in order to find the behavior of the deflection angle in the strong deflection limit, we follow the method developed by Bozza \cite{Bozza}. We first define two new variables $y$ and $z$:
\begin{eqnarray}
 y&=&A(r),\\
 z&=&\frac{y-y_{0}}{1-y_{0}},
\end{eqnarray}
where $y_{0}=A_{0}$. With the two new variables, the total azimuthal angle (\ref{azimuthal11}) can be expressed as
\begin{eqnarray}
 \phi_{f}(r_{0})=\int_{0}^{1}R(z,r_{0})f(z,r_{0})dz, \label{phiintegral}
\end{eqnarray}
where the functions $R(z,r_{0})$ and $f(z,r_{0})$ are defined as
\begin{eqnarray}
 R(z,r_{0})&=&\frac{2\beta(1-y_{0})}{A'}\frac{\sqrt{B|A_{0}|}(D+2\beta LA)}
                  {\sqrt{4AC^{2}+CD^{2}}},\label{Rz0}\\
 f(z,r_{0})&=&\frac{1}{\sqrt{\frac{\text{sgn}(A_{0})}{C}
                        \big[CA_{0}-AC_{0}+\beta
                           L(AD_{0}-A_{0}D)\big]}}.\label{fz}
\end{eqnarray}
These metric coefficients without the subscript ``0" are evaluated at $r=A^{-1}\big((1-y_{0})z+y_{0}\big)$. Note that the function $R(z,r_{0})$ is regular for all values of $z$ and $r_{0}$, while $f(z,r_{0})$ diverges at $z=0$. Therefore, we are motivated to separate the integral (\ref{phiintegral}) into two parts:
\begin{eqnarray}
 \phi_{f}(r_{0})=\phi_{f}^{R}(r_{0})+\phi_{f}^{D}(r_{0}),
\end{eqnarray}
where the regular part $\phi_{f}^{R}(r_{0})$ and the divergent part $\phi_{f}^{D}(r_{0})$ are given by
\begin{eqnarray}
 \phi_{f}^{R}(r_{0})&=&\int_{0}^{1}g(z,r_{0})dz,\\
 \phi_{f}^{D}(r_{0})&=&\int_{0}^{1}R(0,r_{\text{c}})f_{0}(z,r_{0})dz,
\end{eqnarray}
with $g(z,r_{0})=R(z,r_{0})f(z,r_{0})-R(0,r_{\text{c}})f_{0}(z,r_{0})$. In order to find the divergence of the integrand, one needs to expand the argument of the square root of $f(z,r_{0})$ to second order at $z=0$, and then the function $f_{0}(z,r_{0})$ reads
\begin{eqnarray}
 f_{0}(z,r_{0})=\frac{1}{\sqrt{p z+q z^{2}+\mathcal{O}(z^{3})}},
\end{eqnarray}
where
\begin{eqnarray}
 p&=&\text{sgn}(A_{0})\frac{(1-A_{0})}{A'_{0}C_{0}}
         \bigg(A_{0}C'_{0}-A'_{0}C_{0}+\beta L(A'_{0}D_{0}-A_{0}D'_{0})\bigg),\\
 q&=&\text{sgn}(A_{0})\frac{(1-A_{0})^{2}}{2C_{0}^{2}A'^{3}_{0}}
       \bigg(2C_{0}C'_{0}A'^{2}_{0}+(C_{0}C''_{0}-2C'^{2}_{0})A_{0}A'_{0}
             -C_{0}C'_{0}A_{0}A''_{0}\nonumber\\
          &&+\beta L\big[A_{0}C_{0}(A''_{0}D'_{0}-A'_{0}D''_{0})
              +2A'_{0}C'_{0}(A_{0}D'_{0}-A'_{0}D_{0})\big]\bigg).
\end{eqnarray}
When $p\neq 0$, the leading order of the divergence in $f_{0}(z, r_{0})$ is $z^{-1/2}$, which can be integrated to give a finite result. However, when $p=0$ and $q\neq 0$, the leading order of the divergence is $z^{-1}$, and the integral will diverge. As a result, $p=0$ can be used to define the photon circle equation. It is worth noting that the photon circle equation defined by $p=0$ and the one defined by the effective potential $V_{\text{eff}}$ (\ref{veff}) are the same. As $r_{0}\rightarrow r_{\text{c}}$, the two expansion coefficients read
\begin{eqnarray}
 p_{\text{c}}&=&0,\\
 q_{\text{c}}&=&\text{sgn}[A_{\text{c}}]\frac{(1-A_{\text{c}})^{2}}
                   {2C_{\text{c}}A_{\text{c}}'^{2}}
       \bigg[A_{\text{c}}C_{\text{c}}''-A_{\text{c}}''C_{\text{c}}
         +\beta L_{\text{c}}(A_{\text{c}}''D_{\text{c}}-A_{\text{c}}D_{\text{c}}'')\bigg],
\end{eqnarray}
where the subscript ``c" means that these metric coefficients are evaluated at $r=r_{\text{c}}$. For the case $r_{0}\sim r_{\text{c}}$, the deflection angle can be expanded in the following form \cite{Bozza}
\begin{eqnarray}
 \alpha(u)=-\bar{a}\ln\big(\frac{u}{u_{\text{c}}}-1\big)
                +\bar{b}+\mathcal{O}(u-u_{\text{c}}),\label{Atheta}
\end{eqnarray}
where all the strong deflection coefficients $\bar{a}$, $\bar{b}$, and $u_{\text{c}}$ depend on the metric coefficients and are evaluated numerically at $r_{\text{c}}$.
The parameter $u=\theta D_{\text{OL}}$, with $\theta$ the angular separation between the lens and the image, and $D_{\text{OL}}$ the observer-lens distance. These strong deflection limit coefficients appearing in (\ref{Atheta}) are
\begin{eqnarray}
 u_{\text{c}}&=&L\big|_{r_{0}=r_{\text{c}}},\\
 \bar{a}&=&\frac{R(0,r_{\text{c}})}{2\sqrt{q_{\text{c}}}}
            =\beta\sqrt{\frac{2A_{\text{c}}B_{\text{c}}}
                   {A_{\text{c}}C_{\text{c}}''-A_{\text{c}}''C_{\text{c}}+
                   \beta u_{\text{c}}(A_{\text{c}}''D_{\text{c}}
                       -A_{\text{c}}D_{\text{c}}'')}},\\
 \bar{b}&=&-\pi\beta+b_{\text{R}}+\bar{a}\ln\bigg(\frac{4q_{\text{c}}C_{\text{c}}}
             {u_{\text{c}}|A_{\text{c}}|(D_{\text{c}}
                      +2\beta u_{\text{c}}A_{\text{c}})}\bigg).
\end{eqnarray}
The coefficient $b_{\text{R}}$ is the regular integral evaluated at the point $r_{\text{c}}$:
\begin{eqnarray}
 b_{\text{R}}=\phi_{\text{R}}(r_{\text{c}})
      =\int_{0}^{1}g(z,r_{\text{c}})dz.
\end{eqnarray}
The coefficient $b_{\text{R}}$ could not be obtained analytically and we will give a numerical result for it.

\begin{figure}
\includegraphics[width=8cm]{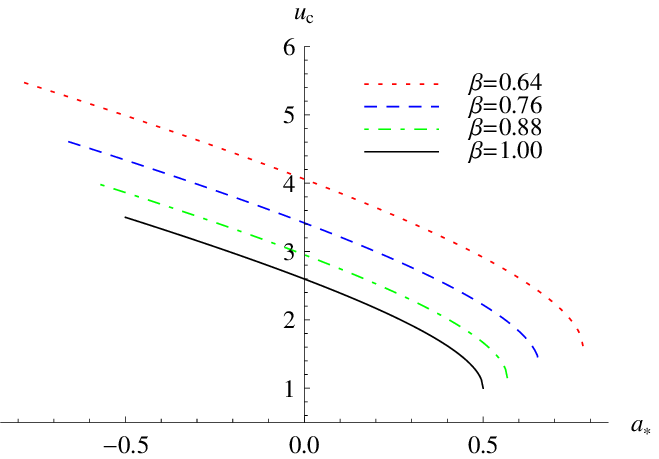}
\includegraphics[width=8cm]{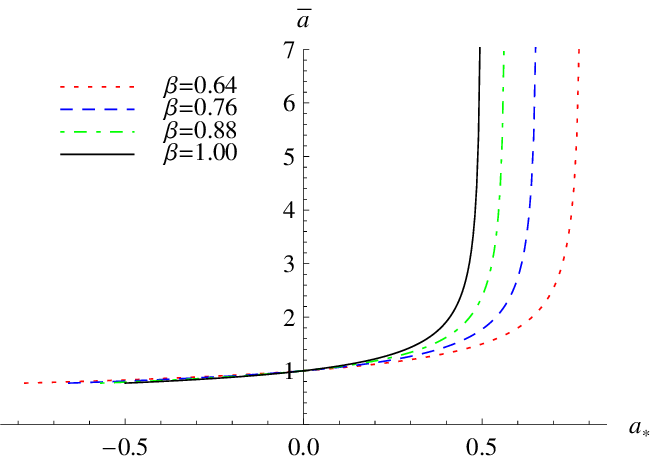}\\ \vspace{0.5cm}
\includegraphics[width=8cm]{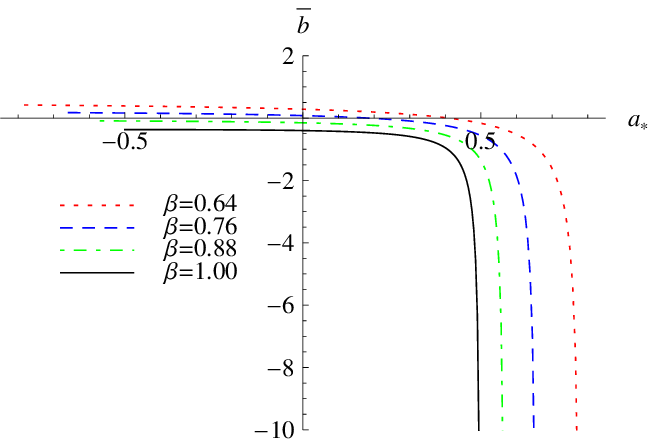}\\
\caption{Strong deflection limit coefficients as a function of the dimensionless spin $a_{*}=a/2M_{\text{phys}}$ for different values of the cosmic string parameter $\beta$.} \label{coefficients1}
\end{figure}

We plot the strong deflection coefficients $u_{\text{c}}$, $\bar{a}$, and $\bar{b}$ as a function of the dimensionless spin $a_{*}$ for different values of the cosmic string parameter $\beta$ in Fig. \ref{coefficients1}. It is clear that the three parameters all depend on the cosmic string parameter $\beta$. The minimum impact parameter $u_{\text{c}}$ has a behavior similar to the circular radius shown in Fig. \ref{RH}. For fixed $\beta$, the coefficient $\bar{b}$ decreases with the dimensionless spin $a_{*}$ while $\bar{a}$ grows with it. When the dimensionless spin $a_{*}$ approaches to $1/(2\beta)$ (extremal black hole case), we find $u_{\text{c}}$ takes a finite value, while $\bar{a}$ goes to positive infinity and $\bar{b}$ goes to negative infinity. In particular, it is worth to pointing out that the divergence of these coefficients implies that the deflection angle in the strong deflection limit no longer represents a reliable description in the regime of high dimensionless spin $a_{*}$. Note that results for the Kerr black hole can be recovered by setting $\beta=1$, and $\bar{b}$ for the Kerr black hole is negative for all values of $a_{*}$. However, for $\beta\neq 1$, it can take positive values for some cases. For fixed $a_{*}$, we find that $u_{\text{c}}$ and $\bar{b}$ decrease as $\beta$ increases, while $\bar{a}$ has a complicated behavior. For fixed $a_{*}$ with a small value, it decreases with $\beta$,
and the result for high $a_{*}$ is the reverse.

\begin{figure}
\includegraphics[width=8cm]{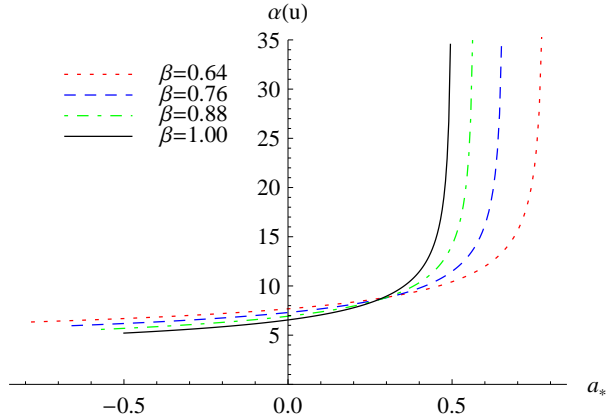}
\caption{Behavior of the deflection angle $\alpha(u)$ as a function of dimensionless spin $a_{*}=a/2M_{\text{phys}}$ with $u=u_{\text{c}}+0.0025$.} \label{Angle}
\end{figure}

With the knowledge of these strong deflection limit coefficients, we can obtain the behavior of the deflection angle $\alpha(u)$ given by (\ref{Atheta}). In Fig. \ref{Angle}, we plot the deflection angle against the black hole dimensionless spin $a_{*}$ with fixed $u=u_{\text{c}}+0.0025$. For each value of the cosmic string parameter $\beta$, we always have a monotonically increasing curve. The deflection goes to infinity when the dimensionless spin $a_{*}$ approaches its maximum value $1/(2\beta)$, which is results from the divergence of the coefficients $\bar{a}$ and $\bar{b}$. It is easy to find that direct photons always have larger deflection angle than retrograde ones. Our result is consistent with \cite{Hackmann}, where the effect of the cosmic string on the light deflection was investigated in detail. We also find that for fixed dimensionless spin $a_{*}$, the behavior is similar to $\bar{a}$.

\subsection{Lens equation in the equatorial plane}
\label{lensequation}

In the previous subsection, we have expressed the deflection angle $\alpha(u)$ as a function of the strong deflection limit coefficients. Numerical results for the coefficients and the deflection angle are analyzed. For the purpose of providing a better description of the geometric disposition of the lens, source, and observer, several lens equations were introduced. Among these equations, Ref. \cite{Bozza103} gave a detailed comparison of each lens equation and suggested that the Ohanian lens equation is the best approximate lens equation. Here, we will give a brief introduction to the lens equation. In this geometric description, one important and necessary ingredient is the optical axis, which is defined as the line joining the observer and the lens. Setting the black hole in the origin, we denote the angle between the direction of the source and the optical axis by $\gamma$. Then the case in which the source, lens, and observer are in perfect alignment corresponds to $\gamma=0$. From the lensing geometry, the angle $\gamma$ can be expressed as
\begin{eqnarray}
 \gamma&=&-\alpha(\theta)+\theta+\overline{\theta}\;\text{mod}\; 2\pi\beta,\label{equatorialequation}\\
 \overline{\theta}&\simeq& \frac{u}{D_{\text{LS}}}
                  \simeq \frac{\theta D_{\text{OL}}}{D_{\text{LS}}}.
\end{eqnarray}
Here, $D_{\text{LS}}$ denotes the distance between the lens and the source, and $D_{\text{OS}}$ measures the distance between the observer and the source. In fact, all these distances are defined in the associated Minkowski space along the optical axis, and they satisfy the relation $D_{\text{OS}}=D_{\text{OL}}+D_{\text{LS}}$. It is also worth noting that these distances are not the true distances between the points of source, lens, and observer. The true distances should be defined as the radial coordinates measured by the original curved metric. However, as pointed out by \cite{Bozza103},  if the source is very close to the optical axis, these distances measured in the Minkowski metric and the original curved metric are approximately equal.

Note that the Ohanian lens equation (\ref{equatorialequation}) is different from the original one \cite{Ohanian} since the azimuth direction of an angle is measured by $\beta\phi$. Multiplying by $1/\beta$ on both sides of the equation, we will obtain the Ohanian lens equation measured in terms of $\phi$. In fact, these lens equations are consistent with each other. The lens equation can be reexpressed as
\begin{eqnarray}
 \gamma=\frac{D_{\text{OL}}+D_{\text{LS}}}{D_{\text{LS}}}\theta-\alpha(\theta)
                  \;\text{mod}\;2\pi\beta.\label{lensgamma}
\end{eqnarray}
Here the angle $\gamma$ can take any value in the interval $[-\pi\beta,\;\pi\beta]$. So, the source and observer could be on opposite sides of the lens, $\gamma=0$, or on the same side, $\gamma=\pi\beta$.

Next, we would like to obtain the positions of the relativistic images by solving the lens equation. For $\theta=u/D_{\text{OL}}\ll 1$, we have $\gamma\simeq-\alpha(\theta)$. With the help of the deflection angle (\ref{Atheta}), one may find
\begin{eqnarray}
 \theta_{n}^{0}=\frac{u_{\text{c}}}{D_{\text{OL}}}(1+e_{n}),
\end{eqnarray}
with $e_{n}=e^{(\gamma+\bar{b}-2n\pi\beta)/\bar{a}}$ and $n$ the number of loops performed by the photon around the black hole. Expanding the deflection angle $\alpha(\theta)$ around $\theta_{n}^{0}$, we have
\begin{eqnarray}
 \alpha(\theta)&=&\alpha(\theta_{n}^{0})
                +\frac{\partial\alpha}{\partial\theta}\big|_{\theta_{n}^{0}}
                  (\theta-\theta_{n}^{0})
                  +\mathcal{O}(\theta-\theta_{n}^{0})\;\text{mod}\;2\pi\beta\nonumber\\
               &\simeq& -\gamma-\frac{\bar{a}D_{\text{OL}}}{u_{\text{c}}e_{n}}
                  (\theta-\theta_{n}^{0})
                    +\mathcal{O}(\theta-\theta_{n}^{0})\;\text{mod}\;2\pi\beta.
\end{eqnarray}
Neglecting the higher order terms and plunging this result into the equatorial lens equation (\ref{lensgamma}), we get the position of the $n$-th relativistic images,
\begin{eqnarray}
 \theta_{n}\simeq\theta_{n}^{0}\bigg(1-
                \frac{u_{\text{c}}e_{n}(D_{\text{OL}}+D_{\text{LS}})}
                  {\bar{a}D_{\text{OL}}D_{\text{LS}}}\bigg).
\end{eqnarray}
Since $D_{\text{OL}}\sim D_{\text{LS}}\gg 1$, we find the correction is much smaller than $\theta_{n}^{0}$. Here we define the north as the direction of the spin. Then, according to the past oriented light ray that starts from the observer and finishes at the source, one may get the following patterns: the resulting images are situated on the eastern side of the black hole for direct photons and on the western side of the black hole for retrograde photons.

\section{Quasi-equatorial lensing by a Kerr black hole pierced by a cosmic string}
\label{Quasiequatiorial}

As shown above, to investigate equatorial lensing, a one-dimensional lens equation is sufficient. However, for further study on the caustic structures, and magnification of images for a quasiequatorial lensing, one needs a two-dimensional lens equation, because the polar angle $\vartheta$, or equivalently the declination $\psi=\pi/2-\vartheta$, should be included in. Here, we only consider the small $\psi$ case for simplicity. As pointed out in \cite{Bozza67,Gyulchev75}, for the rotating black hole, there is a precession of the photon orbit.

\subsection{Precession of the orbit at small declinations}
\label{Precession}

\begin{figure}
 \includegraphics[width=8cm]{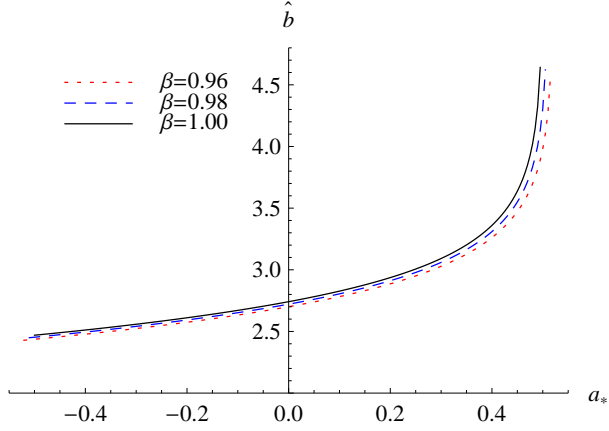}
\caption{The coefficient $\hat{b}$ vs the dimensionless spin $a_{*}=a/2M_{\text{phys}}$ for different values of the cosmic string parameter $\beta$.} \label{AhatBhat}
\end{figure}

\begin{figure}
 \includegraphics[width=8cm]{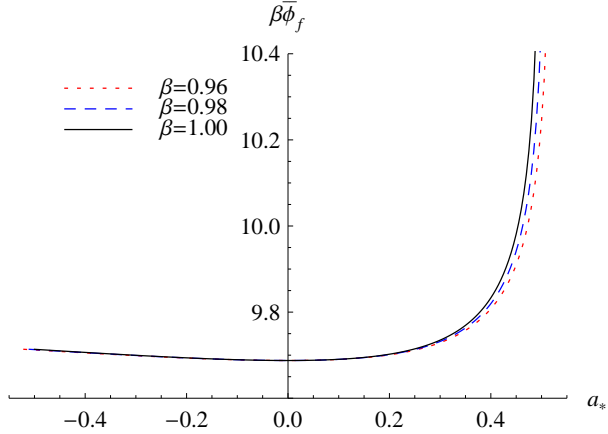}
\caption{Variation of the value of $\beta\bar{\phi}_{f}$ with the dimensionless spin $a_{*}=a/2M_{\text{phys}}$ for fixed $u=u_{\text{c}}+0.0025$.} \label{Phif}
\end{figure}

In this subsection, we would like to study the precession of the orbit at small declinations. Thus, we restrict ourselves to light rays with small inward inclination $\psi_{0}$ and small height $h$ compared to the projected impact parameter $u$. Then one may have $\psi_{0}\sim h/u$. From Eqs. (\ref{angular}) and (\ref{cater}), we have, retaining the first relevant terms,
\begin{eqnarray}
 L&\simeq&u,\\
 \mathcal{K}&\simeq&h^{2}+\bar{u}^{2}\psi_{0}^{2},
\end{eqnarray}
with the new parameters defined as
\begin{eqnarray}
 \bar{u}^{2}&=&u^{2}-a_{*}^{2}.
\end{eqnarray}
With the help of (\ref{Theta}) and (\ref{phit}), we obtain a simple evolution equation for $\psi$ as a function of the azimuthal angle $\beta\phi$ for the Kerr black hole pierced by a cosmic string,
\begin{eqnarray}
 \frac{d\psi}{\beta d\phi}=\mp\omega(\beta\phi)\sqrt{\bar{\psi}^{2}-\psi^{2}},\label{psiphi}
\end{eqnarray}
where
\begin{eqnarray}
 \bar{\psi}&=&\sqrt{\frac{h^{2}+\bar{u}^{2}\psi_{0}^{2}}{\bar{u}^{2}}},\label{phibar}\\
 \omega(\beta\phi)&=&\sqrt{L^{2}-a_{*}^{2}}\frac{\beta a_{*}^{2}+r(\beta r-1)}
                  {[a+ L(\beta r-1)]r}. \label{omega}
\end{eqnarray}
When $\beta=1$, $\omega(\beta\phi)$ recovers the result for the Kerr black hole, and when $a_{*}=0$, we always have $\omega(\beta\phi)=1$. The solution of (\ref{psiphi}) is
\begin{eqnarray}
 \psi(\beta\phi)=\bar{\psi}\cos(\beta\bar{\phi}+\beta\phi_{0}), \label{phi0}
\end{eqnarray}
with $\beta\bar{\phi}=\int^{\phi}_{0}\omega(\beta\phi')\beta d\phi'$ and $\beta\phi_{0}$ a constant. Considering the photon coming from infinity and returning to infinity, we have the deflection angle for this case,
\begin{eqnarray}
 \beta\bar{\phi}_{f}=\int_{0}^{\beta\phi_{f}}\omega(\beta\phi')\beta d\phi'.
\end{eqnarray}
Further, the integral can be written as
\begin{eqnarray}
 \beta\bar{\phi}_{f}=2\int_{r_{0}}^{\infty}\omega(r)\frac{\beta d\phi}{dr}dr
               =\int_{0}^{1}R_{\omega}(z,r_{0})f(z,r_{0})dz,
\end{eqnarray}
with
\begin{eqnarray}
 R_{\omega}(z,r_{0})=\omega(r)R(z,r_{0}),
\end{eqnarray}
where the quantities $R(z, r_{0})$ and $f(z, r_{0})$ are given by Eqs. (\ref{Rz0}) and (\ref{fz}), respectively. Since $\omega(r)$ adds no singularities, it can be absorbed into the regular function $R(z, r_{0})$; therefore, we can apply the same technique used in Sec. \ref{Deflectionequatorial} to this case. Thus, we obtain
\begin{eqnarray}
 \beta\bar{\phi}_{f}&=&-\hat{a}\ln\bigg(\frac{u}{u_{\text{c}}}-1\bigg)+\hat{b},\label{phiangle}\\
 \hat{a}&=&\frac{R_{\omega}(0,r_{\text{c}})}{2\sqrt{q_{\text{c}}}}=1,\\
 \hat{b}&=&\hat{b}_{R}
       +\hat{a}\ln\bigg(\frac{4q_{\text{c}}C_{\text{c}}}
           {u_{\text{c}}|A_{\text{c}}|(D_{\text{c}}+2\beta u_{\text{c}} A_{\text{c}})}\bigg),
\end{eqnarray}
where
\begin{eqnarray}
 \hat{b}_{R}=\int_{0}^{1}
      \bigg[R_{\omega}(z,r_{0})f(z,r_{0})
    -R_{\omega}(0,r_{\text{c}})f_{0}(z,r_{0})\bigg]dz.
\end{eqnarray}
For different values of the cosmic string parameter $\beta$ and the dimensionless spin $a_{*}$, we always have $\hat{a}=1$, while the coefficient $\hat{b}$ depends on $\beta$, which is depicted in Fig. \ref{AhatBhat}. We also find that, for fixed $\beta$, $\hat{b}$ grows with $a_{*}$ and diverges for the extremal black hole case. For fixed $a_{*}$, it increases with the cosmic string parameter $\beta$. Variation of $\beta\bar{\phi}_{f}$ with the dimensionless spin $a_{*}$ is presented in Fig. \ref{Phif}. From this figure, we can see that $\beta\bar{\phi}_{f}$ has a behavior similar to the deflection angle $\alpha(u)$ shown in Fig. \ref{Angle}.


With the detailed analysis of the $\omega(\beta\phi)$, it is easy to find that, for the positive dimensionless spin $a_{*}$, we have $\omega<1$, which leads to $\beta\bar{\phi}_{f}<\beta\phi_{f}$. Thus, the orbital plane suffers a counterclockwise precession which means that after each loop an additional $\beta\Delta\phi$ is necessary to reach the same declination $\psi$, while for the negative $a_{*}$ yields the opposite result.

The integration constant $\beta\phi_{0}$ in Eq. (\ref{phi0}) can be fixed by the initial condition. Taking the same initial condition as \cite{Bozza67}, we have
\begin{eqnarray}
 \beta\phi_{0}=-\text{sgn}(h)\arccos\left(-\frac{\psi_{0}}{\psi}\right).\label{phi02}
\end{eqnarray}
The declination of the outward photon is
\begin{eqnarray}
 \psi_{f}\equiv\psi(\beta\phi_{f})=\bar{\psi}\cos(\beta\bar{\phi}_{f}+\beta\phi_{0}).\label{psi02}
\end{eqnarray}
Or, with the help of (\ref{phi02}), we can express $\psi_{f}$ as
\begin{eqnarray}
 \psi_{f}=-\psi_{0}\cos\beta\bar{\phi}_{f}-\frac{h}{\bar{u}}\sin\beta\bar{\phi}_{f}.
         \label{psif}
\end{eqnarray}

\subsection{Lensing at small declinations}
\label{smalldeclinations}

It is clear that the deflection angles for the equatorial lensing and quasiequatorial lensing are different from each other. So, it is natural that the lens geometries for both cases are different. In fact, in order to study the positions and magnification of the images at small declination, besides the equatorial lens equation (\ref{equatorialequation}), the polar lens equation is also needed. In this subsection, we will study it at small declination. Here we set the source at height $h_{S}$ above the equatorial plane and the observer at height $h_{O}$. During the whole trajectory of the photon, we assume that the following relation holds:
\begin{eqnarray}
 (h_{O},h_{S})\ll u\ll (D_{OL},D_{LS}),
\end{eqnarray}
which ensures the small declination condition. For the incoming and outgoing light rays, we have the simple geometric relations
\begin{eqnarray}
 h=h_{O}+D_{OL}\psi_{0},\label{h}\\
 h_{S}=h_{f}+D_{LS}\psi_{f}.\label{hs}
\end{eqnarray}
Next, we would like to determine the inclination $\psi_{0}$. By symmetry between the incoming parameter and the outgoing parameter, we can express the parameter $\bar{\psi}$, in terms of $\psi_{f}$ and $h_{f}$, as
\begin{eqnarray}
 \bar{\psi}=\sqrt{\frac{h_{f}^{2}+\bar{u}^{2}\psi_{f}^{2}}{\bar{u}^{2}}}.
\end{eqnarray}
With the help of (\ref{phibar}), (\ref{phi02}) and (\ref{psi02}), we have
\begin{eqnarray}
 h_{f}=\bar{u}\bar{\psi}\sin(\beta\bar{\phi}_{f}+\beta\phi_{0})
      =-\bar{u}\psi_{0}S-hC,\label{hf}
\end{eqnarray}
where $S=\sin\beta\bar{\phi}_{f}$ and $C=\cos\beta\bar{\phi}_{f}$. Substituting (\ref{psif}) and (\ref{hf}) into (\ref{hs}), and discarding higher order terms, we can obtain the quantity $h_{S}$. Finally, substituting $h$ from (\ref{h}) into $h_{S}$, we obtain the lens equation in the polar direction,
\begin{eqnarray}
 h_{S}=h_{O}\bigg(\frac{D_{\text{LS}}}{\bar{u}}S-C\bigg)-\bigg((D_{\text{OL}}+D_{\text{LS}})C
    -\frac{D_{\text{OL}}D_{\text{LS}}}{\bar{u}}S-\bar{u}S\bigg)\psi_{0}.
\label{polar}
\end{eqnarray}
The inclination $\psi_{0}$ can be solved through this equation, which is related to the heights of the observer and the source,
\begin{eqnarray}
 \psi_{0,n}=\frac{h_{S}\bar{u}-D_{\text{LS}}h_{O}S_{n}
                +\bar{u}h_{O}C_{n}}
            {\bar{u}^{2}S_{n}-\bar{u}(D_{\text{LS}}+D_{\text{OL}})C_{n}
             +D_{\text{OL}}D_{\text{LS}}S_{n}},\label{phi0n}
\end{eqnarray}
where $S_{n}$ and $C_{n}$ are the values of $S$ and $C$ evaluated at $\beta\bar{\phi}_{f}=\beta\bar{\phi}_{f,n}$. With a straightforward calculation, we can get the height $h_{n}$ from Eq. (\ref{h}), which has the same denominator as that of (\ref{phi0n}). Notice that our approximation above requires the two constraints $\psi_{0}\ll 1$ and $h\ll u$. Since the last term in the denominator of (\ref{phi0n}) dominates, one obtain approximately $\psi_{0,n}\sim -h_{O}/D_{\text{OL}}$ for a general $\beta\bar{\phi}_{f}$. So, both the constraints are satisfied for $\psi_{0,n}$. However, in the neighborhood of $\beta\bar{\phi}_{f}=k\pi\beta$, the denominator of $h_{n}$ can be very small \cite{Bozza67}, which will destroy the small declination constraint. Thus, their denominators determine the positions of the caustic points, i.e.,
\begin{eqnarray}
 K(\gamma)=-\bar{u}(D_{\text{LS}}+D_{\text{OL}})C_{n}
             +D_{\text{OL}}D_{\text{LS}}S_{n},\label{K}
\end{eqnarray}
where we have neglected the high term in $\bar{u}$. Solving $K(\gamma)=0$, we can obtain the caustic points.

\section{Magnification and caustic points}
\label{Magnification}

This section is devoted to the study of the magnification and caustic points of black hole lensing. Let us start with the magnification, which is defined as the ratio of the angular area element of the image and the corresponding angular area element of the source seen by the observer without the lens. The two angular areas are
\begin{eqnarray}
 d^{2}\mathcal{A}_{I}&=&d\theta d\psi_{0},\\
 d^{2}\mathcal{A}_{S}&=&\frac{D_{\text{LS}}d\gamma dh_{S}}
                             {(D_{\text{OL}}+D_{\text{LS}})^{2}}.
\end{eqnarray}
From the equatorial lens equation (\ref{equatorialequation}), retaining the dominant terms, we get
\begin{eqnarray}
 \frac{\partial \gamma}{\partial \theta}\simeq
         -\frac{\bar{a}D_{\text{OL}}}{u_{\text{c}}e_{\gamma}},
\end{eqnarray}
where $e_{\gamma}=e^{(\bar{b}+\gamma)/\bar{a}}$. Note that the number of loops made by the photon can be coded in $\gamma$. Thus, $\gamma$ can take an arbitrary negative value and $\gamma$ mod $2\pi\beta$ denotes the angular position of the source. The number of loops performed by the photon is $n=\frac{\pi\beta-\gamma}{2\pi\beta}$. It is also worth noting that different values of $\gamma$, differing by a multiple of $2\pi\beta$ represent the same source position with respect to the lens. In particular, $\gamma=2n\pi\beta$ corresponds to the case in which the source is located behind the lens, while $\gamma=2(n+1)\pi\beta$ corresponds to the one before the lens. From the polar lens equation (\ref{polar}), we obtain
\begin{eqnarray}
 \frac{\partial h_{S}}{\partial\psi_{0}}=\frac{K(\gamma)}{\bar{u}}.
\end{eqnarray}
With these quantities, we arrive at the magnification
\begin{eqnarray}
 \mu=\frac{d^{2}\mathcal{A}_{I}}{d^{2}\mathcal{A}_{S}}
    =\frac{(D_{\text{OL}}+D_{\text{LS}})^{2}}{D_{\text{OL}}D_{\text{LS}}}
      \frac{\bar{u}u_{\text{c}}e_{\gamma}}{\bar{a}K(\gamma)}.
\end{eqnarray}
It is clear that, at the caustic points $K(\gamma)=0$, we will get a divergent magnification $\mu$, and the corresponding images are called the enhanced images. At the lowest order in $u/D_{\text{OL}}$, the solution $K(\gamma)=0$ is
\begin{eqnarray}
 \beta\bar{\phi}_{f}\simeq k\pi\beta.
\end{eqnarray}
Using (\ref{Atheta}) and (\ref{phiangle}), together with $\gamma=-\alpha(\theta)$, we get
\begin{eqnarray}
 -\left(\frac{\gamma+\bar{b}}{\bar{a}}\right)\hat{a}+\hat{b}=k\pi\beta.
\end{eqnarray}
The solution is
\begin{eqnarray}
 \gamma_{k}=-\bar{b}+\frac{\bar{a}}{\hat{a}}(\hat{b}-k\pi\beta).
\end{eqnarray}
For each $k$, there is one caustic point for direct photons and one for retrograde photons. The case $k=1$ describes the weak field caustic point, and $k\geq 2$ is for the strong field limit approximation.

\begin{figure}
 \includegraphics[width=8cm]{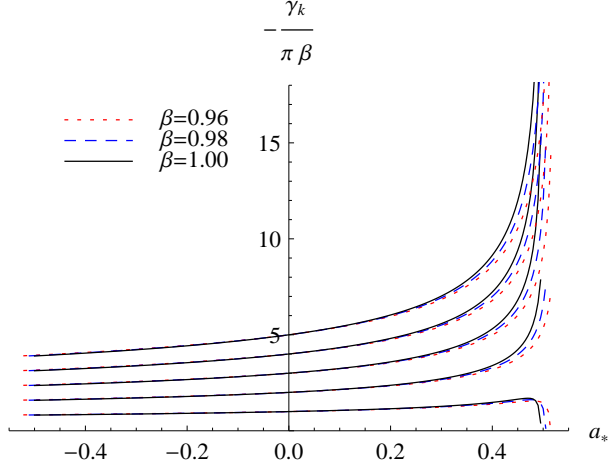}
\caption{The angular position of the first five relativistic caustic points. $k=2,3,4,5,6$ from bottom to top.} \label{Gamma}
\end{figure}

\begin{figure}
 \includegraphics[width=8cm]{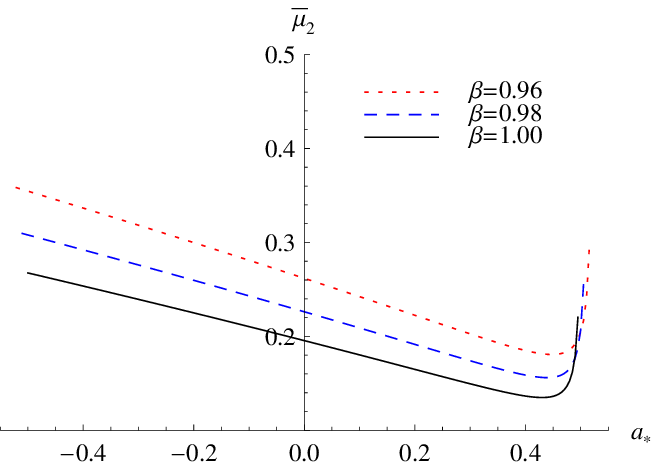}
 \includegraphics[width=8cm]{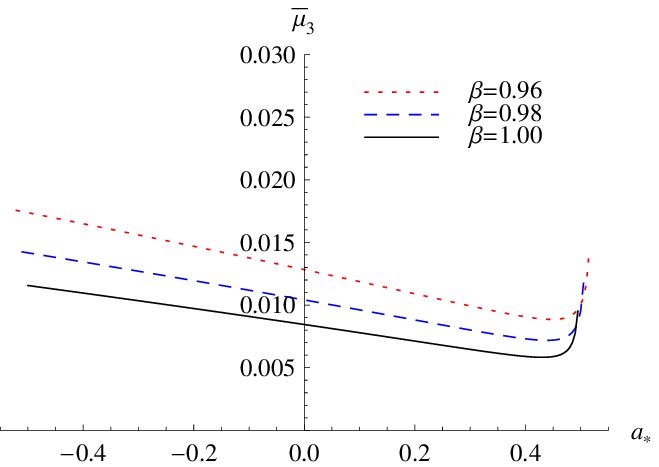}\\ \vspace{0.5cm}
 \includegraphics[width=8cm]{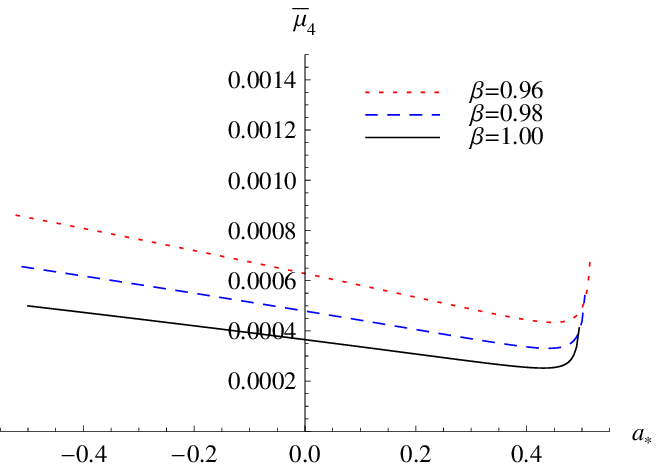}
 \includegraphics[width=8cm]{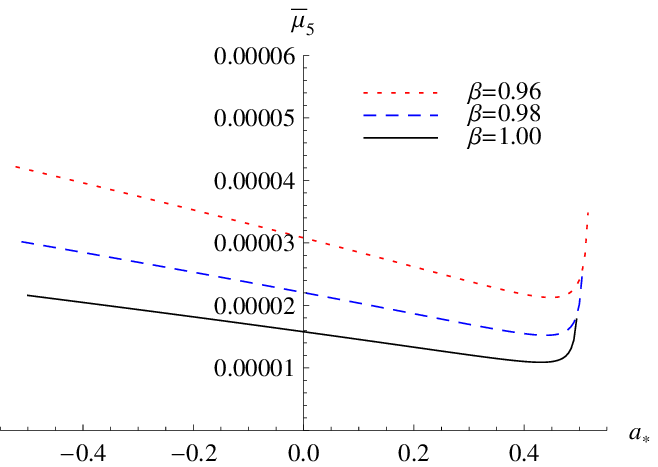}
\caption{Behavior of the magnifying power $\bar{\mu}_{k}$ for $k=2,3,4,5$.} \label{Mu}
\end{figure}

In Fig. \ref{Gamma}, the positions of the first five relativistic caustic points are plotted against the dimensionless spin $a_{*}$ for different values of the cosmic string parameter. When $\beta=1$, the results for the Kerr black hole will be reproduced, and when $\beta=1$ and $a_{*}=0$, the results for the Schwarzschild black hole will be recovered. In the Schwarzschild black hole case, we have $\gamma_{k}=-(k-1)\pi$. Therefore, all the caustic points for the Schwarzschild black hole are located at the optical axis. However, for the Kerr black hole with or without cosmic string, all caustic points of the source are not on the optical axis. They are anticipated for negative dimensionless spin $a_{*}$ and delayed for positive $a_{*}$. It is obvious that the caustic curves can move very far from the optical axis at large value of the dimensionless spin $a_{*}$. Regarding $2\pi\beta$ as a Riemann fold, we can see that the caustic points drift so much that they can even change several Riemann fold. It is also worth noting that the cosmic string parameter $\beta$ has a stronger impact on direct photons than retrograde ones.
In order to describe the magnification for the enhanced images created by a source close to the caustic points, we expand (\ref{K}) around $\gamma_{k}$ and retain the first term,
\begin{eqnarray}
 K(\gamma)=-\frac{\hat{a}D_{\text{OL}}D_{\text{LS}}}{\bar{a}}
               (\gamma-\gamma_{k}(a_{*})).
\end{eqnarray}
Thus, we can express the magnification of the enhanced images in the form
\begin{eqnarray}
 u_{k}^{\text{enh}}&=&\frac{(D_{\text{OL}}+D_{\text{LS}})^{2}}
                {D_{\text{OL}}^{2}D_{\text{LS}}^{2}}
                \frac{\bar{\mu}_{k}}{|\gamma-\gamma_{k}|},\\
 \bar{\mu}_{k}&=&\bar{u}u_{\text{c}}e_{\gamma_{k}}\hat{a}^{-1},\label{mu}\\
 e_{\gamma_{k}}&=&e^{(\hat{b}-k\pi\beta)/\hat{a}}.
\end{eqnarray}
The quantity $\bar{\mu}_{k}$ in fact denotes the magnifying power for the Kerr black hole pierced by a cosmic string close to the caustic points. We plot the numerical results for the magnifying power $\bar{u}_{k}$ in Fig. \ref{Mu}. From this, we can see that $\bar{u}_{k}$ decreases with the dimensionless spin $a_{*}$ for fixed $\beta$, and decreases with $\beta$ for fixed $a_{*}$. For the Kerr black hole with or without a cosmic string, we can see that the magnifying power diverges at $a_{*}=1/(2\beta)$. However, we should keep in mind that the behavior of the magnifying
power around $a_{*}=1/(2\beta)$ is inaccurate because the strong field limit approximation breaks down. Furthermore, the magnification falls rapidly with the increase of $k$.

From (\ref{mu}), we can see that the shape of $\bar{\mu}_{k}$ remains more or less the same for different numbers of $k$. And for the two numbers $k$ and $k+1$, we have
\begin{eqnarray}
 \frac{\bar{\mu}_{k+1}}{\bar{\mu}_{k}}\simeq e^{-\pi\beta}.\label{mumu}
\end{eqnarray}
When $\beta=1$, this describes the Kerr black hole case. It is also clear that this result is independent of the dimensionless spin $a_{*}$ and the Komar's mass $M_{\text{phys}}$. Thus, by measuring the value of $\frac{\bar{\mu}_{k+1}}{\bar{\mu}_{k}}$ from astronomical observations, we are allowed to determine the value of $\beta$.

\section{Critical curves and caustic structure}
\label{Critical}

For the nonrotating black hole lens, if the source, lens, and observer are strictly aligned, one will get a large Einstein ring and two infinite series of concentric relativistic Einstein rings, very close to the minimum impact angle $\theta_{\infty}$ \cite{Bozza}. However, for the rotating black holes \cite{Bozza67,Gyulchev75}, the result changes. The nontrivial caustic structures will appear, and the caustics will drift away from the optical axis and acquire finite extension. For a black hole with high spin, only one image will be observed rather than two infinite series of relativistic images.

Here we would like to discuss the critical curves and caustic structures of the Kerr black hole pierced by a cosmic string. The effects resulting from $\beta$ and $a_{*}$ will also be analyzed in the following. For the Kerr black hole with a cosmic string, the intersections of the critical curve with the equatorial plane are given by
\begin{eqnarray}
 \theta_{k}^{\text{cr}}\simeq\theta_{k}^{\text{0,cr}}
                   \bigg(1-\frac{u_{\text{c}}e_{\gamma_{k}}
                  (D_{\text{OL}}+D_{\text{LS}})}
                   {\bar{a}D_{\text{OL}}D_{\text{LS}}}\bigg),
\end{eqnarray}
with $\theta_{k}^{\text{0,cr}}=\frac{u_{\text{c}}}{D_{\text{OL}}}(1+e_{\gamma_{k}})$. Suppose that the gravitational field of the supermassive black hole at the center of our Milky Way can be described by the metric (\ref{metric}). The mass of the supermassive black hole is estimated to be $M=2.8\times 10^{6}M_{\odot}$, and $D_{OL}=8.5$ kpc, $D_{\text{LS}}=1.0$ kpc. Then the numerical results for $\theta_{k}^{\text{cr}}$ are listed in Table \ref{Table}. From these results, we can see that the cosmic string parameter $\beta$ has a weak influence on the critical points. It is also clear that the critical points are close to the optical axis $\theta=0$ for the positive dimensionless spin $a_{*}$ and farther away for the negative one. So one could conclude that the critical curves are distorted and shifted towards the western side.

\begin{table}[h]
\begin{center}
\begin{tabular}{c c c c c c}
  \hline\hline
    & & $\theta_{2}^{\text{cr}}$($\mu$arcsec) & $\theta_{3}^{\text{cr}}$($\mu$arcsec) & $\theta_{4}^{\text{cr}}$($\mu$arcsec) & $\theta_{5}^{\text{cr}}$($\mu$arcsec) \\\hline
           & $\beta=0.96$\quad & 20.0836 & 19.4840 & 19.4546 & 19.4531 \\
  $a_{*}=-0.15$& $\beta=0.98$\quad & 19.6492 & 19.1187 & 19.0943 & 19.0931 \\
           & $\beta=1.00$\quad & 19.2380 & 18.7687 & 18.7484 & 18.7475 \\\hline
           & $\beta=0.96$\quad & 18.2039 & 17.6064 & 17.5771 & 17.5757 \\
  $a_{*}=0.00$ & $\beta=0.98$\quad & 17.7710 & 17.2424 & 17.2181 & 17.2170 \\
           & $\beta=1.00$\quad & 17.3612 & 16.8937 & 16.8735 & 16.8726 \\\hline
           & $\beta=0.96$\quad & 16.1640 & 15.5626 & 15.5332 & 15.5317 \\
  $a_{*}=0.15$ & $\beta=0.98$\quad & 15.7284 & 15.1962 & 15.1718 & 15.1706 \\
           & $\beta=1.00$\quad & 15.3160 & 14.8451 & 14.8248 & 14.8239 \\
  \hline\hline
\end{tabular}
\\
\caption{Intersections of the critical curves with the equatorial plane in the strong deflection limit.}
\label{Table}
\end{center}
\end{table}

\begin{figure}
 \includegraphics[width=8cm]{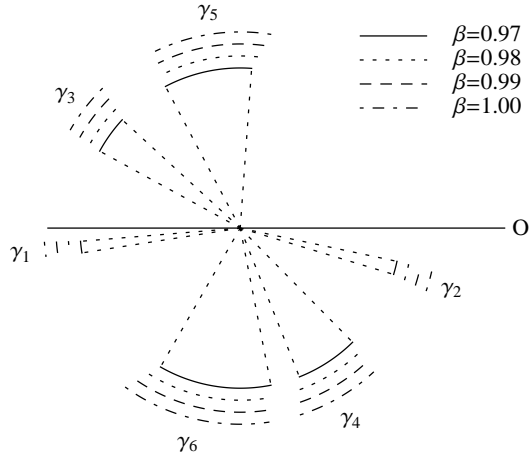}
\caption{The first six caustics for the Kerr black hole pierced by a cosmic string with fixed dimensionless spin $a_{*}=0.15$, marked by the arcs between $\gamma_{k}(-|a_{*}|)/\beta$ and $\gamma_{k}(|a_{*}|)/\beta$.} \label{Caustic1}
\end{figure}

\begin{figure}
 \includegraphics[width=8cm]{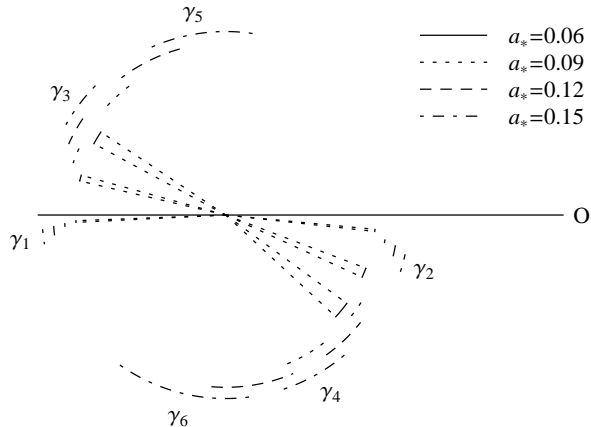}
\caption{The first six caustics for the black hole for different values of the dimensionless spin $a_{*}=a/2M_{\text{phys}}$ with $\beta=0.98$.} \label{Caustic2}
\end{figure}

As noted above, for a rotating black hole, caustics are nonvanishing extensions. Recall that $\gamma$ is measured by $\beta\phi$; thus, mapping the intersections of the $k$-th caustic with the equatorial plane into a flat plane, it will be determined by $\gamma_{k}(-|a_{*}|)/\beta$ and $\gamma_{k}(|a_{*}|)/\beta$. In the flat plane, the first six caustics are shown in Fig. \ref{Caustic1} with fixed dimensionless spin $a_{*}$ and in Fig. \ref{Caustic2} with fixed $\beta$, seen from the northern direction. The nonrelativistic caustic $\gamma_{1}$ stays close to the optical axis, and the relativistic ones drift in the clockwise direction. Relativistic caustics $\gamma_{2}$, $\gamma_{4}$, and $\gamma_{6}$ are shifted toward the western side, and $\gamma_{3}$, $\gamma_{5}$ are shifted toward the eastern side. It is clear that, as $k$ increases, the caustics get larger and farther from their initial position on the optical axis. From Fig. \ref{Caustic1}, we can see that, for fixed dimensionless spin $a_{*}$, the larger value of $\beta$ is, the farther the caustics shift. For fixed $\beta$ shown in Fig. \ref{Caustic2}, the caustics are shifted much farther with the dimensionless spin $a_{*}$.

\section{Discussions and summary}
\label{Summary}

In this paper, we have studied the equatorial and quasiequatorial gravitational lensings by the Kerr black hole pierced by a cosmic string in the strong deflection limit. Supposing that the massive compact object at the center of our Galaxy can be described by a Kerr black hole pierced by a cosmic string, we compute the strong deflection limit coefficients and the deflection angle. The photon circle, deflection angle, and strong deflection limit coefficients are found to depend on $\beta$. When comparing these parameters to these of the Kerr black hole, we find that there is a significant cosmic string parameter $\beta$ effect on these lensing observables.

When $a_{*}=0$, the lensing by the Schwarzschild black hole with or without a cosmic string can be recovered. For a nonrotating static spherically symmetric black hole, if the point source is perfectly aligned along the optical axis, it will produce infinitely bright images. However, extremely faint relativistic images will be produced if the source is not aligned with the optical axis. The caustics for the black hole are points located at the optical axis in front of or behind the black hole lens. So the relativistic images are maximally amplified altogether when the source lies on the optical axis.

For Kerr black hole lensing (the case of $\beta=1$ and $a_{*}\neq 0$), the phenomenon will be quiet different. Its caustics drift away from the optical axis. And if the source is close to one of the caustic points, then it cannot be close to any others; this leads to the result that only one image at a time can be enhanced, while others will be quite faint. Supposing the source is aligned with the caustic point $\gamma_{k}(|a_{*}|)$, then the outermost relativistic image on the eastern side will be enhanced. If the source is set at $\gamma_{k}(-|a_{*}|)$, then only the one on the western side is enhanced. If the source is placed between $\gamma_{k}(|a_{*}|)$ and $\gamma_{k}(-|a_{*}|)$, we will obtain two enhanced images.

For black hole lensing due to the Kerr black hole pierced by a cosmic string, we have $\beta\neq 1$ and $a_{*}\neq 0$. The photons have a bigger impact parameter for small $\beta$, which implies that the photons are captured more easily by the black hole with larger $\beta$. Meanwhile, the cosmic string parameter $\beta$ indeed has an effect on the other quantities. Here, we list some of them:

(1) the deflection angle $\alpha(u)$ decreases with $\beta$ for fixed dimensionless spin $a_{*}$ and impact parameter $u$ (this can also be found in \cite{Hackmann});

(2) the two-dimensional lens equation obviously depends on $\beta$;

(3) the intersections of the critical curve with the equatorial plane $\theta_{k}^{\text{cr}}$ decrease with $\beta$;

(4) the caustic gets larger when $\beta$ increases;

(5) the magnifying power $\bar{\mu}_{k}$ depends on the cosmic string parameter $\beta$ (see Fig. \ref{Mu});

(6) the caustics shift farther away from the optical axis for small value of $\beta$.\\
Comparing these results with the Kerr black hole without a cosmic string, we come to the conclusion that there is a significant effect of $\beta$ on the observable parameters. Furthermore, as we analyzed above, for a fixed cosmic parameter $\beta$, the dimensionless spin $a_{*}$ also has an important influence on black hole lensing. In particular, from (\ref{mumu}), we can see that the quantity $\frac{\bar{\mu}_{k+1}}{\bar{\mu}_{k}}$ is independent of the black hole spin and mass. This result may provide us with a possible way to determine the value of the cosmic string parameter $\beta$ by measuring $\frac{\bar{\mu}_{k+1}}{\bar{\mu}_{k}}$ from astronomical observations in the near future.

\section*{Acknowledgement}
The authors would like to thank the anonymous referees whose comments largely helped us in improving the original manuscript. S.-W. Wei would like to thank Professor J.-X. Lu at the Interdisciplinary Center for Theoretical Study (ICTS), University of Science and Technology of China (USTC) for warm hospitality. The authors also wish to thank Professor Songbai Chen for useful discussions and valuable comments. This work was supported by the Program for New Century Excellent Talents in University, the Huo Ying-Dong Education Foundation of the Chinese Ministry of Education (No. 121106), the Fundamental Research Funds for the Central Universities (No. lzujbky-2012-k30), and the National Natural Science Foundation of China (No. 11075065).

\end{document}